\documentclass[floatfix,aps,prl,superscriptaddress,amsmath,groupedaddress,twocolumn]{revtex4}

\usepackage{color,hangcaption,hhline,psfrag,rotating,amssymb}
\usepackage[hang,nooneline]{subfigure}
\usepackage{dcolumn}
\usepackage{hyperref}
\usepackage{graphicx}% Include figure files

\newcommand{\nn}{\nonumber\\}
\newcommand{\ds}{\displaystyle}
\newcommand{\ident}{{\bf 1}} 
\newcommand{\be}{\begin{equation}}
\newcommand{\ee}{\end{equation}}
\newcommand{\beq}{\begin{eqnarray}}
\newcommand{\eeq}{\end{eqnarray}}

\newcommand{\imag}{\mbox{i}}

\begin{document}
\title{The $\Delta(1232)$ axial charge and  form factors from lattice QCD}
\author{Constantia Alexandrou}
\affiliation{Department of Physics, University of Cyprus, P.O. Box 20357,
        1678 Nicosia, Cyprus}
\affiliation{Computation-based Science and Technology Research Center, The Cyprus Institute, P.O. Box 27456, 1645 Nicosia, Cyprus}
\author{Eric B. Gregory}
\affiliation{Department of Physics, University of Cyprus, P.O. Box 20357,
        1678 Nicosia, Cyprus}
\author{Tomasz Korzec}
\affiliation{Institut f\"ur Physik, Humboldt Universit\"at zu Berlin, 
Newtonstrasse 15, 12489 Berlin,Germany}
\author{Giannis Koutsou}
\affiliation{Computation-based Science \& Technology Research Center, The Cyprus Institute, P.O. Box 27456, 1645 Nicosia, Cyprus}
%\affiliation{J\"ulich Supercomputing Center, Forschungszentrum J\"ulich, 
%D-52425, J\"ulich, Germany}
% \affiliation{Bergische Universit\"at Wuppertal, Gaussstr. 20, D-42119, 
%Wuppertal, Germany}
\author{John W. Negele}
  \affiliation{Center for Theoretical Physics, 
Laboratory for Nuclear Science and 
        Department of Physics, Massachusetts Institute of
        Technology, Cambridge, Massachusetts 02139, U.S.A.}
\author{Toru Sato}
  \affiliation{Department of Physics, Osaka University, Osaka 560-0043, Japan}
\author{Antonios Tsapalis}     
\affiliation{Hellenic Naval Academy, Hatzikyriakou Ave., Pireaus 18539, 
Greece}       
\affiliation{Department of Physics, National Technical University of
       Athens, Zografou Campus 15780, Athens, Greece}

%\thanks{}
%\affiliation{}


\date{\today}

\begin{abstract}
We present the first calculation on the $\Delta$ axial-vector  
and pseudoscalar form factors using lattice QCD. 
Two Goldberger-Treiman relations are derived and examined.
A combined chiral fit is performed to the nucleon axial charge, N to $\Delta$ axial
transition coupling constant and $\Delta$ axial charge.  
\end{abstract}

%\pacs{}
%\keywords{}

\maketitle

%\section{Introduction}

{\it Introduction.} Numerical solution of Quantum Chromodynamics (QCD), the underlying theory of the 
strong interactions, has proved a very successful approach in providing
a theoretical understanding of baryon structure. During the 
last few years, simulations of the discretized theory known as lattice
QCD have
included dynamical quarks with near to physical mass values~\cite{Jansen:2008vs}.
A success of these recent simulations is  the calculation of the low-lying
hadron spectrum~\cite{Durr:2008zz,Alexandrou:2009qu,Aoki:2009ix} showing
  agreement with the experimental values. 
The lattice set-up can also be applied  to compute
quantities that are not known experimentally. In this work we report
the first calculation of the $\Delta$ axial-vector and pseudoscalar
form factors (FFs).

Understanding the structure of the  $\Delta$ resonance  has great relevance to 
nuclear phenomenology. The  $\Delta $ is a rather broad resonance close
to the $\pi N$ threshold. It therefore 
couples strongly to nucleons and pions making it an important
ingredient in chiral expansions~\cite{Bernard:2005fy, Hemmert:1997ye,Jenkins:1991es,Fettes:2000bb}.
The $\Delta $ baryon resists experimental probing due to its short lifetime
($\sim 10^{-23}$ s)~\cite{Kotulla:2002cg,LopezCastro:2000cv}.
Its axial charge
and $\pi$-$\Delta$ coupling constants
that are needed as input in chiral Lagrangians are difficult to measure.
Baryon chiral expansion calculations that include the $\Delta$ explicitly
follow one of two  strategies as far as the determination
of these parameters is concerned. The first is to relegate the 
axial charge  to one of many fit parameters, and fit 
using  lattice \cite{Bernard:2005fy,Bernard:2007zu},
experimental \cite{Jenkins:1991es},
 or partial-wave calculation data \cite{Fettes:2000bb}.
The second is to use estimates based on phenomenology such as the relation  
between the nucleon axial 
charge $g_A$ that is well measured and the $\Delta$  axial charge, which
can be  derived from the large-$N_c$ limit
\cite{Dashen:1993jt} or $SU(4)$ symmetry
\cite{Brown:1975di}.
The Goldberger-Treiman relation is then used to get the effective $\pi \Delta \Delta$ coupling.

Quite recently, groups have calculated $\Delta$ axial charge
 through QCD sum rules
\cite{Choi:2010ty} and $\chi$PT \cite{Jiang:2008we}, and both have noted the 
lack of an explicit lattice calculation of this quantity.
First-principles lattice QCD calculations can probe the structure of the $\Delta$ 
and indeed recent studies have produced calculations of the  $\pi N \Delta$ 
coupling \cite{Alexandrou:2007xj,Alexandrou:2010uk} and the electromagnetic 
form-factors of the  $\Delta$ \cite{Alexandrou:2009hs}. 
Using our lattice QCD formulation, we are then  well-positioned to 
calculate the axial charge of the 
and the effective pion-$\Delta$ coupling,  
$G_{\pi\Delta\Delta}$, as well as examine 
the Goldberger-Treiman relations as a
way to relate the $\Delta$ axial charge to $G_{\pi\Delta\Delta}$.

In this letter we present the first lattice calculation of the axial-vector and pseudoscalar
form factors of the $\Delta$ baryon. %The dominant axial form factor leads 
%
%directly to a calculation of $g_{\Delta\Delta}$, 
%the axial charge of the $\Delta$. The dominant pseudoscalar form-factor
%yields the effective pion-$\Delta$ coupling constant, $G_{\pi\Delta\Delta}$. 
At non-zero momentum transfer $q^2$, we find a second pseudoscalar form-factor, 
yielding a second effective coupling constant, $H_{\pi\Delta\Delta}$. 
These calculations lead to two Goldberger-Treiman relations, which are indeed 
satisfied by the lattice results.

%\section{Decomposition}
{\it Axial-vector and Pseudoscalar Matrix Element.}
We consider the  matrix element of a current X   between $\Delta^+$ states  
\begin{equation}
\label{ax_decomp_min}
\langle \Delta(p_f,s_f) | X| \Delta(p_i,s_i)\rangle = 
\overline{u}_\sigma(p_f,s_f)
\left[{\mathcal O}^{X}\right]^{ \sigma\tau} 
u_\tau(p_i,s_i), \nonumber
\end{equation}
where $u_\sigma$ denotes the Rarita-Schwinger vector-spinor 
($\sigma,\tau$ are Lorentz indices), and $p_i$ and $ p_f$ are the initial and
final momenta of the $\Delta$.
For the axial-vector current
$A_\mu^a(x)= \overline{\psi}(x)\gamma_\mu\gamma_5 \frac{\tau^a}{2}\psi(x)$
the matrix element 
can be written 
 in terms of four Lorentz-invariant FFs, labeled $g_1$, $g_3$, $h_1$ and $h_3$:
\begin{eqnarray}
\label{ax_operator}
\left[{\mathcal O}^{\rm A_\mu^3}\right]^{ \sigma\tau} &=&
-\frac{1}{2}\left[
g^{\sigma\tau}
\left(g_1(q^2)\gamma_\mu\gamma^5 
    + g_3(q^2) \frac{q_\mu}{2m_\Delta}\gamma^5\right)\right.\nonumber\\
&&\hspace*{-0.6cm} + \left.\frac{\ds q^\sigma q^\tau}{\ds 4m_\Delta^2}
\left(h_1(q^2)\gamma_\mu\gamma^5 
   + h_3(q^2) \frac{q_\mu}{2m_\Delta}\gamma^5\right)\right],
\end{eqnarray}
where $q=p_f-p_i$ is the momentum transfer.
For
the pseudoscalar current 
$
P^a(x)= \overline{\psi}(x)\gamma_5 \frac{\tau^a}{2}\psi(x)
$, the matrix element  can be written in terms of two FFs, to be defined below
in relation to the Goldberger-Treiman relations.
%\begin{equation}
%\label{ps_decomp_min}
%\langle \Delta(p_f,s_f) | P| \Delta(p_i,s_i)\rangle = 
%\overline{u}_\sigma(p_f,s_f)
%\left[{\mathcal O}^{ {\rm PS }}\right]^{ \sigma\tau} 
%u_\tau(p_i,s_i)\quad,\nonumber
%\end{equation}
%in terms of two pseudoscalar form factors, $\tilde{g}$ and $\tilde{h}$:
%\begin{equation}
%\left[{\mathcal O}^{ {\rm PS }}\right]^{ \sigma\tau} =
%\frac{1}{2}\left[
%-g^{\sigma\tau}
%\tilde{g}(q^2)\gamma^5 
%+\frac{\ds q^\sigma q^\tau}{\ds 4m_\Delta^2}
%\tilde{h}(q^2)\gamma^5 \right]\quad.
%\end{equation}
%The normalization of Eqs.~(\ref{ax_decomp_min},\ref{ps_decomp_min}) is 
%clearly isospin dependent. 

{\it Lattice Evaluation.}
Axial form factors on the lattice are extracted in a standard way from the 
three-point function
\beq 
\langle G_{\sigma \tau}^{\Delta A_\mu^3 \Delta} (t_f, t ;
{\bf
p}_f, {\bf p}_i; \Gamma_\rho) \rangle  = 
\sum_{{\bf x}_2, \;{\bf
x}_1} e^{-i {\bf p}_f \cdot {\bf x}_2 } e^{+i {\bf q}
\cdot {\bf x}_1 } \; \Gamma_\rho^{\beta \alpha} \nn 
\langle \Omega | 
T\left[\chi_{\Delta}^{\sigma \alpha} ({\bf x}_2,t_2) A_\mu^3({\bf
x}_1,t_1) \bar{\chi}_{\Delta}^{\tau \beta} ({\bf 0},0) \right] |\Omega
\rangle  
\label{3pt} 
\eeq
where  $\bar{\chi}_\Delta$ is an interpolating 
operator creating a state with the quantum numbers of the
$\Delta^+$~\cite{Alexandrou:2009hs} and
$\Gamma_\rho$ is a set of projectors, given by $\Gamma_4 = \frac{1}{4}(\ident + \gamma_4)$ and 
$\Gamma_k = \imag \Gamma_4 \gamma_5 \gamma_k$. 
A similar three-point function with $A_\mu^3 \rightarrow P^3$ 
is required for the extraction of the pseudoscalar FFs.
Technically these are evaluated via the {\it sequential inversion through
the sink}~\cite{Alexandrou:2009hs} at fixed sink time-slice $t_f$, while the $A_\mu^3$ and $P^3$ 
operator insertion is supplied at all intermediate t-slices 
$(0 \le t\le t_f)$ and Fourier-transformed for all momenta ${\bf q}$ at a small
extra CPU cost.

The kinematics are fixed to a static $\Delta$ sink 
$({\bf p}_f = {\bf 0}, {\bf q} = - {\bf p}_i )$.
%The $\Delta$ two-point function is also required (summation over $k$)
%\beq 
%&&G_{kk}(\Gamma_4,{\bf p}, t) =\nn
%&&\sum_{\bf x}
%e^{-i {\bf p} \cdot {\bf x}} \; \Gamma_4^{\beta \alpha} 
%\langle \Omega | 
%T\left[\chi_{\Delta}^{k \alpha} ({\bf x},t)
% \bar{\chi}_{\Delta}^{k \beta} ({\bf 0},0) \right] |\Omega
%\rangle  
%\label{2pt} 
%\eeq
Denoting for convenience Monte-Carlo averages 
$G_{\sigma \tau}^X (\Gamma_\rho,{\bf q}, t) = 
\langle G_{\sigma \tau}^{\Delta X \Delta} (t_f, t \,;
{\bf 0}, {\bf -q}; \Gamma_\rho) \rangle $ for $X=A_\mu^3$ or $P^3$, we
construct the optimal  ratio of three-point to two- point
functions 
\begin{eqnarray}
R_{\sigma \tau}^{X}(\Gamma_\rho,{\bf q},t) 
&=& \frac{G_{\sigma \tau}^{X}(\Gamma_\rho,{\bf q },t)}
{G_{\Delta}(\Gamma_4,{\bf 0}, t_f)} \nonumber \\
 & & \hspace*{-3cm}\sqrt{\frac{G_{\Delta}(\Gamma_4, {\bf -q}, 
t_f-t)G_{\Delta}(\Gamma_4, {\bf 0}  ,t)G_{\Delta}(\Gamma_4, {\bf 0},t_f)}
  {G_{\Delta}(\Gamma^4, {\bf 0}, t_f-t)G_{\Delta}(\Gamma_4, {\bf -q},t)
G_{\Delta}(\Gamma_4, {\bf -q},t_f)}} \nonumber \\ 
& &\stackrel{\tiny{\begin{array}{c}t_f-t
\rightarrow\infty\\t-t_i \rightarrow\infty\end{array}}}{\longrightarrow} 
\Pi_{\sigma\tau}^X(\Gamma_\rho,{\bf q}) \quad,
\label{ratio}
\end{eqnarray}
where $G_{\Delta}(\Gamma^4, {\bf p}, t)$ is the $\Delta$ propagator of momentum ${\bf p}$.
This ratio 
eliminates unknown field renormalization constants and leading time dependences
and tends to a constant at large Euclidean time separations 
$t_f-t$ and $t$.
A careful optimization in the space 
of  the source-sink Lorentz
indices $\sigma,\tau,\rho$ is required 
and only {\it two} linear combinations of 
sequential sources suffice to
provide all four axial %FFs $g_1,g_3,h_1,h_3$, 
and  two pseudoscalar FFs.%\tilde{g}$ and $\tilde{h}$.
%These
%two linear combinations lead to 
%$\Pi_{\sigma \tau}^X(\Gamma_\rho,{\bf q})$ suffice for the evaluation
%of all 
%\vspace{-0.2cm}
%%\beq
%\label{type12}
%\Pi_I^X ({\bf q})&=& \sum_{\rho=1}^3 \sum_{\sigma=\tau=1}^3 
%\Pi_{\sigma \tau}^X(\Gamma_\rho,{\bf q})\nn
%\Pi_{II}^X ({\bf q})&=& \sum_{\rho=1}^3 \sum_{\sigma=\tau=1}^3 
%\epsilon_{\sigma\tau\rho} \Pi_{\sigma \tau}^X(\Gamma_4,{\bf q})
%\eeq
%where $\epsilon_{\sigma\tau\rho}$ is the fully antisymmetric 3-d tensor.

Smearing techniques are  implemented  resulting in
satisfactory suppression of excited state effects allowing the source-sink distance to be fixed at about 1 fm~\cite{Alexandrou:2009hs}.
Lattice computations of the  matrix elements of the axial-vector and pseudoscalar currents   for all transition momenta
vectors ${\bf q}$ contributing to a given value of $Q^2=-(p_f-p_i)^2$ 
are simultaneously analyzed and the overconstrained system 
determines the form factors through a global $\chi^2$ minimization.
%The full expressions relating the measured plateaus of eq.~\ref{type12} 
%to the form factors will be given elsewhere. 
We note  that
$O(500)$ lattice measurements are involved in the extraction of the form factors for  $Q^2$-values up to $\sim~2$~GeV$^2$.

The parameters of the lattice ensembles 
used in this calculation are given in Table \ref{Table:params}. 
The quenched Wilson
fermions gauge configurations enable the extraction
of the FFs with small statistical errors.
 In addition, we obtained the FFs using 
dynamical domain-wall valence quarks  matched to staggered 
sea fermions~\cite{Bernard:2001av}. For the computation of the
$\Delta$ axial charge we also use $N_f=2+1$ domain wall fermions~\cite{Aoki:2010dy} corresponding to a pion mass of about 300 MeV in order to perform
the chiral
extrapolation.
In all cases the $u$ and $d$ quarks are degenerate whereas the mass of the 
strange quark in the dynamical simulations is set to its physical mass. For the pion
masses of these simulations the $\Delta$ is stable.

\begin{table}[t]
\small
\begin{center}
\begin{tabular}{cccccc}
\hline\hline 
V& stat. & $m_\pi$ (Gev) & $m_N$ (GeV) & $m_\Delta$ (GeV)\\ 
\hline
\multicolumn{5}{c}{Quenched Wilson fermions}\\
\multicolumn{5}{c}{$\beta=6.0,~~a^{-1}=2.14(6)$~GeV} \\
\hline
$32^3\times 64$& 200  &0.563(4)& 1.267(11) & 1.470(15)\\
$32^3\times 64$& 200  &0.490(4)& 1.190(13) & 1.425(16)\\
$32^3\times 64$& 200  &0.411(4)& 1.109(13) & 1.382(19)\\
% &  &$\kappa_c$ =0.1571& 0.& & 0.938(9) &\\
\hline
\hline
\multicolumn{5}{c}{Mixed action, $a^{-1} = 1.58(3)$~GeV} \\
\multicolumn{5}{c}{
Asqtad ($am_{\mbox{\tiny u,d/s}} = 0.02/0.05$),  
DWF ($ am_{\mbox{\tiny u,d}} = 0.0313$)}\\ 
%    $V$ & \# confs & & $m_\pi$  & $m_\pi / m_\rho$ &
%    $m_N$  & $m_\Delta$ \\ 
%     &  & & (GeV) & &
%    (GeV) & (GeV) \\ 
\hline
$20^3\times 64$ &264 &   0.498(3) & 1.261(17)& 1.589(35)\\
\multicolumn{5}{c}{
Asqtad ($am_{\mbox{\tiny u,d/s}} = 0.01/0.05$),  
DWF ($ am_{\mbox{\tiny u,d}} = 0.0138$)}\\ \hline
$28^3\times 64$ &550 &   0.353(2) & 1.191(19)& 1.533(27)\\
\hline\hline
\multicolumn{5}{c}{Domain Wall Fermions (DWF)} \\
%\multicolumn{5}{c}{
% $ m_{\mbox{\tiny u,d}}/m_s=0.005/0.04$,  $a^{-1} =1.73(3)$~GeV}\\
%$24^3\times 64$ & &   0.329(1) &1.130(6) &1.457(11) \\
\multicolumn{5}{c}{
 $ m_{\mbox{\tiny u,d}}/m_s=0.004/0.03$, $a^{-1} =2.34(3)$~GeV}\\\hline
$32^3\times 64$ &1452 &   0.297(5) &1.27(9) &1.455(17) \\
\hline
\hline
\end{tabular}
\end{center}
\caption{Ensembles and parameters used in this work.  We give in the first column  the
lattice size, in the second the statistics, in the third, fourth and fifth the pion, nucleon and $\Delta$ mass in GeV respectively.}
\label{Table:params}
\end{table}

\begin{figure}
\begin{center}
%\hspace{0.02in}
\includegraphics[width=\linewidth,height=1.3\linewidth]{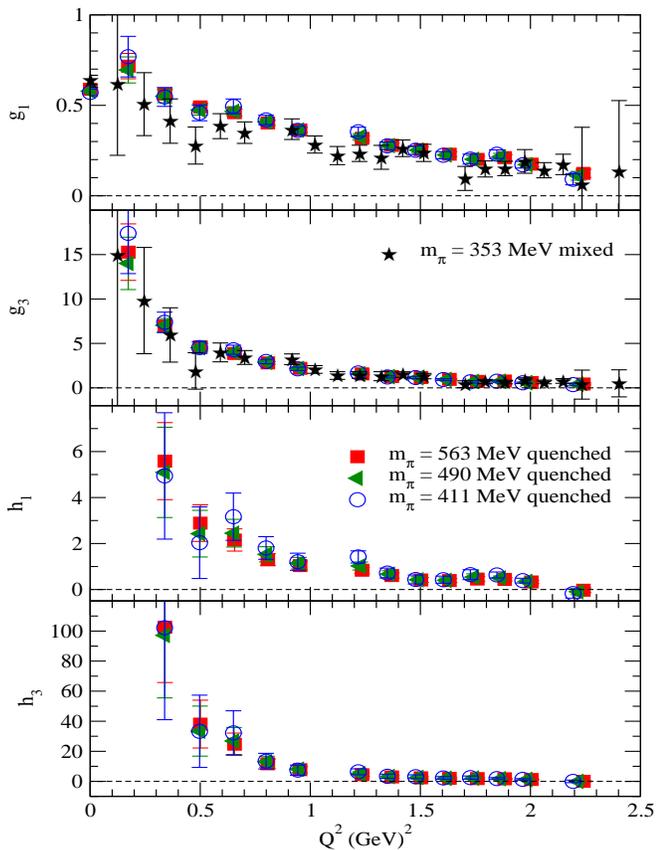}
\end{center}
\caption{Results for the axial form-factors, $g_1$, $g_3$, $h_1$ and $h_3$. Noisier 
mixed-action results are consistent and omitted for clarity for the $h_1$ and $h_3$ FFs.
% Filled squares, triangles and open circles denote results in the
%quenched theory with $m_\pi=0.563\,,0.490\,,0.411$~GeV respectively and asterisks results using the mixed action.}
}
\label{axial_ff_figs}
\end{figure}

%\section{$\Delta$ axial charge}
{\it Lattice results on the $\Delta$ axial form factors:}
In Fig.~\ref{axial_ff_figs} we  show  the 
four axial $\Delta$ form-factors,
$g_1$, $g_3$, $h_1$ and $h_3$, as a function of the momentum transfer 
 $Q^2$. As can be seen, results obtained with the mixed action are in agreement with quenched results for $g_1$ and $g_3$. A similar conclusion holds for $h_1$ and $h_3$ albeit with much larger statistical errors in the case of the mixed action approach that we therefore omit from the plots for clarity. 
%\\{\bf : Fits and dipole mass on $g_1$ ??} \\
%The same analysis
%that yields the form-factors gives us directly the full 
%matrix element $\mathcal{O}^{\mu A}(Q^2=0)$ at $Q^2=0$.
The value of the  matrix element at $Q^2=0$ is connected to the axial 
charge  defined 
 by $\langle\Delta^{++}|A_\mu^3|\Delta^{++}\rangle -\langle\Delta^{-}|A_\mu^3|\Delta^{-}\rangle = G_{\Delta\Delta}{\mathcal M}_\mu$~\cite{Jiang:2008we}. At $q^2=0$ this is
 $G_{\Delta\Delta}=-3g_1(0)$.

{\it Pseudoscalar FFs and the Goldberger-Treiman Relations}.
The $\Delta$ axial charge enters in 
baryon $\chi$PT expressions of many important quantities such as the axial
charge of the nucleon. Many phenomenological results
 rely on this value, which is usually treated as a fit parameter
to be determined  
from fits to experimental or lattice data. It can be related to the 
$\pi\Delta\Delta$ coupling via the Goldberger-Treiman relation.
%There have been several sum-rules calculations of the effective 
%$\pi\Delta\Delta$ coupling~\cite{Belyaev:1984ib,Zhu:2000zd,erkol_thesis}. 
In Ref.~\cite{Jido:1999hd} symmetry arguments in a quartet scheme 
where $N^*_+$, $N^*_-$, $\Delta_{+}$ and $\Delta_{-}$ form a chiral multiplet, 
lead to the
conclusion
 that $\pi\Delta_{\pm}\Delta_{\pm}$ 
couplings (with like-charged $\Delta$s) are forbidden at tree-level. 
Quark-model arguments~\cite{Brown:1975di} suggest that the 
$G_{\pi\Delta\Delta}=(4/5)G_{\pi NN}$. Clearly a 
non-perturbative calculation within lattice QCD  of this coupling,
as presented in this work,
provides  valuable input to phenomenology.

\begin{figure}
\begin{center}
\includegraphics[width=\linewidth,height=0.8\linewidth]{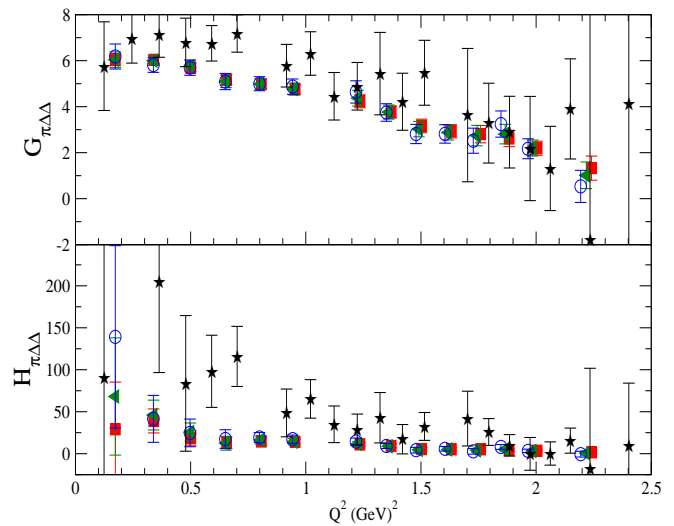}
\end{center}
\caption{The pseudoscalar $\Delta$ form factors $G_{\pi\Delta\Delta}$ and 
$H_{\pi\Delta\Delta}$ for the quenched and dynamical ensembles.
}
\label{eff_couplings}
\end{figure}

Partial Conservation of the Axial Current (PCAC) when applied
 to the hadronic world leads to valuable phenomenological
predictions such as the Goldberger-Treiman (GT) relation, 
originally derived for the nucleon state.
% Lattice QCD has recently studied
%and confirmed the validity of this relation in the nucleon sector. 
Similarly, a non-diagonal GT relation, applicable to the axial $N-\Delta$ transition
is formulated and relates the axial $N-\Delta$ coupling $c_A$ to the 
$\pi N \Delta$ effective coupling. 
%A recent lattice calculation has shown 
%the validity of this relation also, using the domain wall formulation which 
%preserves the chiral symmetry on the lattice. 
PCAC on the hadron level reads:
$
\partial^\mu A_\mu^a=f_\pi m_\pi^2 \pi^a ~.
$
In the SU(2) symmetric limit of QCD with $m_q$ denoting 
the up/down mass, the pseudo-scalar density is related  
to the divergence of the axial-vector current through the
axial Ward-Takahashi identity (AWI) $
 \partial^\mu A_\mu^a = 2 m_q P^a ~.
$
Taking matrix elements of the LHS of the AWI identity in $\Delta$ states 
we can define two
Lorentz-invariant $\pi \Delta \Delta$ form factors,
 $G_{\pi\Delta\Delta}(q^2)$
and $H_{\pi\Delta\Delta}(q^2)$ factoring out the pion pole as dictated by 
PCAC
\beq
\label{eff_couple_eq}
&\langle\Delta (p_f,s_f) |P^3| \Delta (p_i,s_i) \rangle  = 
-\frac{1}{2m_q}\frac{f_\pi m_\pi^2 }{(m_\pi^2-q^2)}\times \nn
&\overline{u}_\sigma \left [g^{\sigma\tau} G_{\pi\Delta\Delta}(q^2)
+\frac{q^\sigma q^\tau}{4m_\Delta^2} H_{\pi\Delta\Delta}(q^2)
\right] \gamma^5 u_\tau ~,
\eeq
%where we define 
%$m_q\tilde{g}(q^2)\equiv
%f_\pi m_\pi^2 G_{\pi\Delta\Delta}(q^2)/(m_\pi^2 - q^2)
%\; ,\;
%m_q\tilde{h}(q^2)\equiv
%f_\pi m_\pi^2 H_{\pi\Delta\Delta}(q^2)/(m_\pi^2 - q^2).
%$
Matrix elements of the AWI identity, 
$ \langle\Delta| \partial_\mu A^\mu | \Delta \rangle = 2m_q\langle\Delta|P | 
\Delta \rangle $ now lead to a matrix equation, satisfied at finite $q^2$,
\beq
\label{GTmatrix}
m_\Delta \left [ g^{\sigma\rho} (g_1 - \tau g_3) 
+\frac{q^\sigma q^\rho}{4m_\Delta^2} (h_1 - \tau h_3) \right ]= \nn
\frac{f_\pi m_\pi^2 }{(m_\pi^2-q^2)}~
\left [g^{\sigma\rho} G_{\pi\Delta\Delta}
+\frac{q^\sigma q^\rho}{4m_\Delta^2} H_{\pi\Delta\Delta}
\right] ~,
\eeq
where $\tau=-q^2/(2m_\Delta)^2$.
We display the pseudoscalar FFs in 
Fig.~\ref{eff_couplings}. The results using dynamical quark simulations 
have increased statistical errors and are consistent with the 
quenched results. The quark mass $m_q$, extracted from the axial 
Ward-Takahashi identity,
and $f_\pi$, calculated from the pion-to-vacuum amplitude, are taken from  
Ref.~\cite{Alexandrou:2007xj}.
%This leads to relations between the axial form factors and 
%the $\pi\Delta\Delta$ form factors,
%\beq
%2 m_\Delta\left(g_1  -\tau g_3\right) &=&\frac{2f_\pi m_\pi^2 
%G_{\pi\Delta\Delta}(q^2) }{m_\pi^2-q^2} ~,\nn
%2 m_\Delta\left(h_1  -\tau h_3\right) &=&\frac{2f_\pi m_\pi^2 
%H_{\pi\Delta\Delta}(q^2) }{m_\pi^2-q^2}~,
%\eeq
%with $\tau=\frac{-q^2}{(2m_\Delta)^2}$. 
%In addition, doting eq.~(\ref{GTmatrix}) with $q_\tau$ leads to the scalar 
%equation
%\beq
%m_\Delta\left[(g_1  -\tau g_3) - \tau (h_1  -\tau h_3)\right] = \nn 
%\frac{f_\pi m_\pi^2 }{m_\pi^2-q^2}~\left[ G_{\pi\Delta\Delta}-\tau
%H_{\pi\Delta\Delta} \right]
%\eeq
%This is a generalized GT-type relation which constrains the pion pole 
%dependence of all form factors. Since $g_1$ and $G_{\pi\Delta\Delta}$ are
%the only finite FFs at the origin, it is easy to see that pion pole dominance
%determines
%\beq
%g_1 &=& \frac{f_\pi}{m_\Delta} G_{\pi\Delta\Delta} \\
%g_3 &=&h_1= \frac{4 f_\pi m_\Delta }{m_\pi^2-q^2} G_{\pi\Delta\Delta}\\ 
%h_3 &=& \frac{16 f_\pi M^3_\Delta }{(m_\pi^2-q^2)^2} G_{\pi\Delta\Delta}\\ 
%H_{\pi\Delta\Delta} &=& 
%\frac{4 M^2_\Delta }{m_\pi^2-q^2} G_{\pi\Delta\Delta} ~.
%\eeq
Assuming pion  pole dominance, which is consistent with our lattice
 results,  that indeed $g_3$ and $h_1$ are of the same order and diverge 
with the pion pole while $h_3$ diverges with (pion pole)$^2$,
% The second 
%pseudoscalar coupling, $H_{\pi\Delta\Delta}$, also diverges with the pion 
%pole. The above information can be contained in 
we obtain a pair of GT-type relations,
valid at finite $q^2$,
\beq
\label{GTR_g}
f_\pi G_{\pi\Delta\Delta}(q^2) = m_\Delta g_1(q^2) \,,  
f_\pi H_{\pi\Delta\Delta}(q^2) = m_\Delta h_1(q^2) \quad.
\eeq
%as well as the equality $g_3 = h_1$.

\begin{figure}[t]
\begin{center}
\includegraphics[width=\linewidth,height=0.8\linewidth]{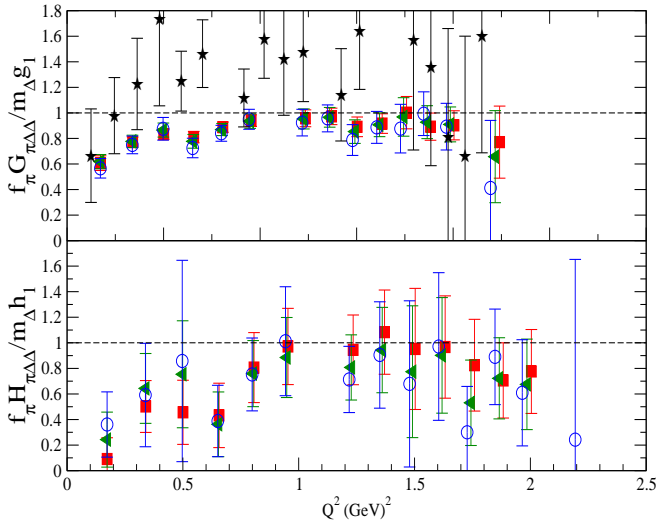}
\end{center}
\caption{The Goldberger-Treiman ratios from Eq.~(\ref{GTR_g}). }
\label{GTR_figs}
\end{figure}

The validity of the GT relations is examined by evaluating
the ratios $f_\pi G_{\pi\Delta\Delta}(Q^2)/m_\Delta g_1(Q^2)$ and
$f_\pi H_{\pi\Delta\Delta}(Q^2)/m_\Delta h_1(Q^2)$ as shown
 in Fig.~\ref{GTR_figs}. For the former ratio for which statistical
errors are smaller the  behavior is similar
to the one obtained for the pseudoscalar nucleon  and
$N-\Delta$ couplings $G_{\pi NN}$ and $G_{\pi N\Delta}$~\cite{Alexandrou:2007xj} for the same ensembles, namely for 
$Q^2\stackrel{>}{\sim}0.8 {\rm GeV}^2$ the lattice data show 
agreement with unity. We expect that the behavior at low $Q^2$ will be 
affected by pion cloud effects as the mass of the pion decreases
towards the physical point.

\begin{figure}
\begin{center}
\includegraphics[width=\linewidth,height=1.2\linewidth]{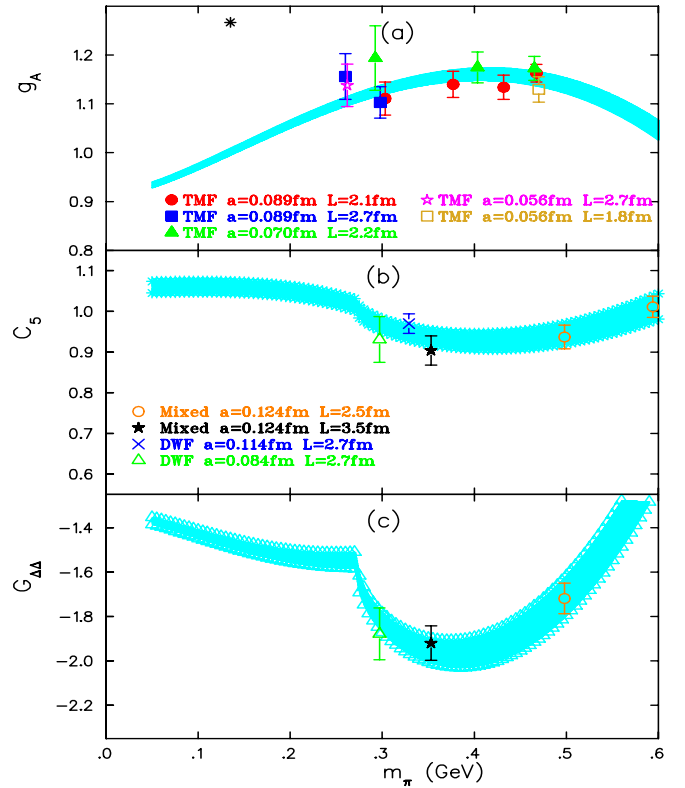}
\end{center}
\caption{Combined chiral fit: (a) Nucleon axial charge, $g_A$, fitted to
 lattice data obtained with $N_f=2$ twisted mass fermions 
(TMF)~\cite{Alexandrou:2010hf}. The physical value is shown by the asterisk;
%(filled circles: a=0.089~fm, L=2.1, filled squares: a=0.089~fm, L=2.8~fm, filled triangles: a=0.070~fm, L=2.2~fm, open square: a=0.056~fm, L=1.8~fm, star: a=0.056~fm, L=2.7~fm); 
(b) Real part of axial N to $\Delta$ transition coupling $C_5(0)$~\cite{Alexandrou:2010uk};
%(open circles: a=0.124~fm, L=2.5~fm, filled square: a=0.124~fm, L=3.5~fm) and dynamical domain wall fermions (cross: a=0.114~fm, L=2.7~fm, open triangle: a=0.084~fm, 2.7~fm); 
(c) Real part of $\Delta$ axial charge
$G_{\Delta \Delta}=-3g_1(0)$.}
\label{axial_charge_figs}
\end{figure}

{\it $\Delta$ axial charge and combined chiral fits.}  Having, for the first
time, a set of  lattice results
 for the axial nucleon charge~\cite{Alexandrou:2010hf},  the axial $N-\Delta$ transition coupling, $C_5$~\cite{Alexandrou:2010uk} and 
the $\Delta$ axial charge, allows us to perform a combined fit to all
three quantities using heavy baryon $\chi$PT in the small scale expansion~\cite{Hemmert:2003cb,Procura:2008ze,Jiang:2008we}.
 The combined fit has seven free parameters, namely the values of the three axial
coupling constants  of the nucleon, the  $N-\Delta$ and the $\Delta$,   three parameters related to the $m_\pi^2$-terms in the chiral expansions  of $g_A$, $C_5$ and $G_{\Delta\Delta}$ and a constant
entering the chiral expression of $C_5$~\cite{Procura:2008ze}.
 As can be seen in Fig.~\ref{axial_charge_figs}, lattice data for all three observables are approximately constant
within the mass range considered.The best fits are shown by the bands that
take into account the statistical errors of the lattice results.
As have been observed in all recent lattice studies, the physical value of 
$g_A$ is underestimated and this combined fit does not provide a possible
resolution to this puzzle.  Having lattice results at pion masses
below 300~MeV will be essential to check the validity of these chiral expansions.

{\it Conclusions.}
We have presented the first calculation of the axial-vector and pseudoscalar 
form factors of the $\Delta$ using lattice QCD.
 From the most general 
decomposition of the axial-vector and pseudoscalar vertex 
we derived two Goldberger-Treiman relations whose validity
is satisfied at the same level of accuracy as that found for the
nucleon case~\cite{Alexandrou:2007xj}.
At zero momentum transfer the $\Delta$ matrix element yields the
 phenomenologically important $\Delta$ axial charge, which in this
work is obtained for pion masses in the
range of about 300~MeV to 500~MeV.
As in the case of the nucleon axial charge, it shows a weak dependence
 on the pion mass in this mass range.
Using lattice results for the axial nucleon charge, the axial N to $\Delta$ transition coupling  and the $\Delta$ axial charge we performed, for the first
time, a combined fit to all three quantities that provides a reasonable description to 
the lattice results. However, these  state-of-the-art lattice results and
chiral perturbation calculations, yield a value for  the nucleon axial charge  lower than its experimental value. Such discrepancies between lattice
and experimental results are seen
in several key hadronic observables~\cite{Alexandrou:2010cm} calling for high accuracy 
lattice calculations  with pion mass below 300~MeV
in order to to gain insight in the chiral
behavior of these fundamental quantities.

{\it Acknowledgments.}
We are grateful to Brian Tiburzi and K. S. Choi for helpful discussions.  EBG was 
supported by Cyprus RPF grant
$\Delta IE\Theta NH\Sigma/\Sigma TOXO\Sigma/0308/07$ and JWN in part by funds provided by the U.S. Department of Energy (DOE) under cooperative research agreement DE-FG02-94ER40818. 
Computer resources were provided by
the National Energy Research Scientific Computing Center supported by the Office of Science of the DOE under Contract No. DE-AC02-05CH11231 and by the J\"ulich Supercomputing Center, awarded under the DEISA Extreme Computing Initiative, co-funded through the EU FP6 project RI-031513 and the FP7 project RI-222919.
This research was in part supported by the Research Executive Agency of the European Union under Grant Agreement number PITN-GA-2009-238353 (ITN STRONGnet).

\end{document}